\documentclass[aps,prl,twocolumn,superscriptaddress]{revtex4-1}
\usepackage{amsfonts}
\usepackage{amsmath}
\usepackage{amssymb}
\usepackage{graphicx}
\usepackage{times}
\usepackage[colorlinks=true,linkcolor=blue,citecolor=red,plainpages=false,pdfpagelabels]
{hyperref}
\usepackage{color}

\providecommand{\U}[1]{\protect\rule{.1in}{.1in}}
\newtheorem{theorem}{Theorem}

\def\Tr{\operatorname{Tr}}

\def\PPT{\operatorname{PPT}}

\def\LOCC{\operatorname{LOCC}}
\def\T{\operatorname{T}}

\begin{document}

\title{Fundamental limits on the capacities of bipartite~quantum~interactions}
\author{Stefan B{\"a}uml}\email{stefanbaeuml@gmx.de}
\affiliation{NTT Basic Research Laboratories, NTT Corporation, 3-1 Morinosato-Wakamiya, Atsugi, Kanagawa 243-0198, Japan}
\affiliation{NTT Research Center for Theoretical Quantum Physics, NTT Corporation, 3-1 Morinosato-Wakamiya, Atsugi, Kanagawa 243-0198, Japan}
\affiliation{QuTech, Delft University of Technology, Lorentzweg 1, 2628 CJ Delft, Netherlands}
\author{Siddhartha Das}\email{sdas21@lsu.edu}
\affiliation{Hearne Institute for Theoretical Physics, Department of Physics and Astronomy, Louisiana State University, Baton Rouge, Louisiana 70803, USA}
\author{Mark M. Wilde}\email{mwilde@lsu.edu}
\affiliation{Hearne Institute for Theoretical Physics, Department of Physics and Astronomy, Louisiana State University, Baton Rouge, Louisiana 70803, USA}
\affiliation{Center for Computation and Technology, Louisiana State University, Baton Rouge, Louisiana 70803, USA}

\date{\today}
\begin{abstract}
Bipartite quantum interactions have applications in a number of different areas of quantum physics, reaching from fundamental areas such as quantum thermodynamics and the theory of quantum measurements to other applications such as quantum computers, quantum key distribution, and other information processing protocols.  A particular aspect of the study of bipartite interactions is concerned with the entanglement that can be created from such interactions. In this paper, we present our work on two basic building blocks of bipartite quantum protocols, namely, the generation of maximally entangled states and secret key via bipartite quantum interactions. In particular, we provide a non-trivial, efficiently computable upper bound on the positive-partial-transpose-assisted (PPT-assisted) quantum capacity of a bipartite quantum interaction. In addition, we provide an upper bound on the secret-key-agreement capacity of a bipartite quantum interaction assisted by local operations and classical communication (LOCC). As an application, we introduce a cryptographic protocol for the read-out of a digital memory device that is  secure against a passive eavesdropper. 
\end{abstract}

\maketitle
\textit{Introduction}---Bipartite quantum interactions are a fundamental feature in numerous areas of quantum physics. Any interaction described by a Hamiltonian of an otherwise closed quantum system with a heat bath realizes a bipartite unitary operation that acts on the quantum system and the bath collectively (cf.~\cite{goold2016role}). Similarly, any noisy evolution or measurement of a quantum system can be described in terms of a bipartite unitary operation acting on the system, as well as an environment or measurement probe system \cite{Sti55,peres2006quantum}. Quantum computation, error correction, and many more information-theoretical applications of quantum physics rely on bipartite unitary quantum operations known as bipartite quantum gates. Examples include the swap gate, the controlled-NOT (CNOT) gate, or  the controlled phase gate~\cite{NC00}.

Going beyond unitary bipartite interactions, one can consider noisy interactions between two quantum systems held by separate parties, Alice and Bob, which can be described by a tripartite unitary operation acting on the two quantum systems as well as an uncorrelated environment, or by a completely positive, trace-preserving map, a bidirectional quantum channel \cite{BHLS03}, acting only on Alice and Bob's systems. Examples of such bidirectional quantum channels are noisy bipartite quantum gates \cite{shor1996fault}, which occur in every realistic implementation of quantum computing, quantum error correction, interactions of two separate quantum systems with a heat bath \cite{goold2016role}, or joint measurements of two quantum systems, as are performed in teleportation or entanglement swapping \cite{BBC+93,zukowski1993event}. 

Depending on the kind of bipartite interaction and the input states, entanglement can be created, destroyed, or changed by the interaction \cite{PV07,HHHH09,huber2015thermodynamic}. Whereas the environment is assumed to be inaccessible to Alice and Bob, it does play a crucial role whenever Alice and Bob are performing bipartite operations in a cryptographic protocol, such as secret key agreement \cite{D05,DW05,HHHO05,HHHO09}. In such a case, it has to be assumed that the eavesdropper can access part of or even the entire environment system. 

In this work, we analyse bipartite interactions in terms of their abilities to create entanglement, as well as secret key. In particular, we focus on determining  bounds on the non-asymptotic quantum and private capacities of bipartite interactions, i.e., the maximum rates at which maximally entangled states or bits of secret key, respectively, can be distilled when a finite number of interactions are allowed. Previous results in this direction include \cite{BRV00,BHLS03,CLL06}, which introduce capacities for classical and quantum communication via bipartite unitary and non-unitary interactions, respectively, as well as a number of results on the entanglement generating capacities or the entangling power of bipartite unitary interactions \cite{ZZF00,LHL03,BHLS03,HL05,LSW09,WSM17,CY16}.

What has been an open question since \cite{BHLS03} is whether there exists a non-trivial, efficiently computable upper bound on the entanglement generating capacity of a bipartite quantum interaction.
 The difficulty in addressing this question is that the protocols for entanglement generation are allowed to use local quantum systems of arbitrarily large dimension, and it might not be clear \textit{a priori} whether such bounds would be possible. Another question left open from prior work is that of considering private communication in the bidirectional context, that is, characterizing the rate at which  secret key bits  can be distilled by Alice and Bob via a bidirectional channel. 

In this paper, we answer the aforementioned questions affirmatively, and our bounds thus serve as benchmarks for assessing the entanglement and secret key agreement capabilities of bipartite interactions. To begin with, we determine an efficiently computable upper bound on the entanglement generating capacity of a bipartite quantum interaction. As examples, we compute this bound for the partial swap operation \cite{audenaert2016entropy}, which is related to how photons interact at a beamsplitter, as well as for
the swap gate concatenated with collective dephasing \cite{palma1996quantum}, which is a kind of bipartite interaction that can occur in a quantum computer. Next, we introduce the secret-key-agreement capacity of a bipartite quantum interaction and provide a general upper bound on it, based on the max-relative entropy of entanglement \cite{D09,Dat09}. Our upper bounds on the quantum and private capacities involve an optimization over bounded quantum systems having a fixed dimension. 

As another contribution, we introduce a cryptographic protocol, which we call private reading, for the read-out of a digital read-only memory device secure against a passive eavesdropper. 
The protocol of private reading is related to quantum reading \cite{BRV00,Pir11}, in which a classical message is sent to a reader, after being stored in a read-only memory device. Physically, the device contains codewords that are sequences of quantum channels, which are chosen from a memory cell (a collection of quantum channels). The information is stored in the choice of channels, and the reader can retrieve the message by using a quantum state to distinguish the channels. In private quantum reading, the message is assumed to be secret, and the reader has to retrieve it in the presence of an eavesdropper.
We determine upper bounds on the performance of any private reading protocol by leveraging the fact that  reading digital information stored in a memory device can be understood as a specific kind of bipartite quantum interaction. 

\textit{Bounds on Quantum and Private Capacities}---Let us begin our discussion of entanglement and secret key distillation via bipartite interactions by defining the relevant entanglement measures and capacities. Let $A'$, $L_A$, and $A$ denote quantum systems held locally by Alice, and let $B'$, $L_B$, and $B$ denote those held  by Bob. Given a bidirectional channel $\mathcal{N}_{A'B'\to AB}$, a completely positive, trace-preserving map from quantum systems $A'B'$  to $AB$, we define the \emph{bidirectional max-Rains information} of $\mathcal{N}$ as $
R_{\max}^{2\to 2}(\mathcal{N}):= \log \Gamma^{2\to2} (\mathcal{N})$, 
where $\Gamma^{2\to2}(\mathcal{N})$ is the solution to the following semi-definite program (SDP):
\begin{align}
\textnormal{minimize}\ &\ \Vert \Tr_{AB}\{V_{L_A ABL_B}+Y_{L_A ABL_B}\}\Vert_\infty\nonumber\\
\textnormal{subject to}\ &\ V_{L_A ABL_B},Y_{L_A ABL_B}\geq 0,\nonumber\\
 \T_{BL_B}&(V_{L_A ABL_B}-Y_{L_A ABL_B})\geq J^\mathcal{N}_{L_AABL_B},
\label{eq:bi-rains-channel-sdp}
\end{align} 
such that $L_A\simeq A'$, and $L_B\simeq B'$. The notation $V_{L_AABL_A},Y_{L_AABL_B}\geq 0$ means that
$V_{L_AABL_A}$ and $Y_{L_AABL_B}$ are constrained to be positive semidefinite operators acting on the Hilbert space of the composite quantum system $L_AABL_B$. Furthermore, the notation 
$L_A\simeq A'$  means that quantum system $L_A$ is isomorphic to system $A'$, which in this case simply means that these systems have the same dimension. Here $\T_X$ denotes the partial transposition with respect to subsystem $X$ and $J^\mathcal{N}:=\mathcal{N}_{A'B'\to AB}(|\Upsilon\rangle\langle\Upsilon|_{L_AL_B:A'B'})$ is the Choi operator of $\mathcal{N}$, with $|\Upsilon\rangle_{L_AL_B:A'B'}:=\sum_{ij}|ij\rangle_{L_AL_B}|ij\rangle_{A'B'}$. The SDP is a generalization of the SDP formulation of the max-Rains information of a point-to-point channel \cite{WFD17}. Whereas $R^{2\to2}_{\max}$ is sufficient to bound entanglement distillation rates, the existence of positive-partial-transpose (PPT) entanglement useful for quantum key distribution \cite{HHHO05,HHHO09} motivates the introduction of a second measure of entanglement, the \textit{bidirectional max-relative entropy of entanglement}:
\begin{equation}\label{eq:bi-max-rel-opt}
E_{\max}^{2\to 2}(\mathcal{N}):= \sup_{\psi_{L_AA'} \otimes \varphi_{B'L_B}}E_{\max}(L_A A; B L_B)_{\mathcal{N}(\psi\otimes\varphi)},
\end{equation} 
where $\psi_{L_AA'}\otimes\varphi_{B'L_B}$ is a pure product state such that $L_A\simeq A'$, and $L_B\simeq B'$ and $E_{\max}(A;B)_{\rho} := \inf\{ \lambda : \rho_{AB} \leq 2^{\lambda} \sigma_{AB} , \sigma_{AB} \in \operatorname{SEP}(A\!:\!B)\}$ denotes the max-relative entropy of entanglement of a state $\rho_{AB}$ \cite{D09,Dat09}, with $\operatorname{SEP}(A\!:\!B)$ denoting the set of all separable states of the bipartite system $AB$. 

\begin{figure}
		\centering
		\includegraphics[width=\linewidth]{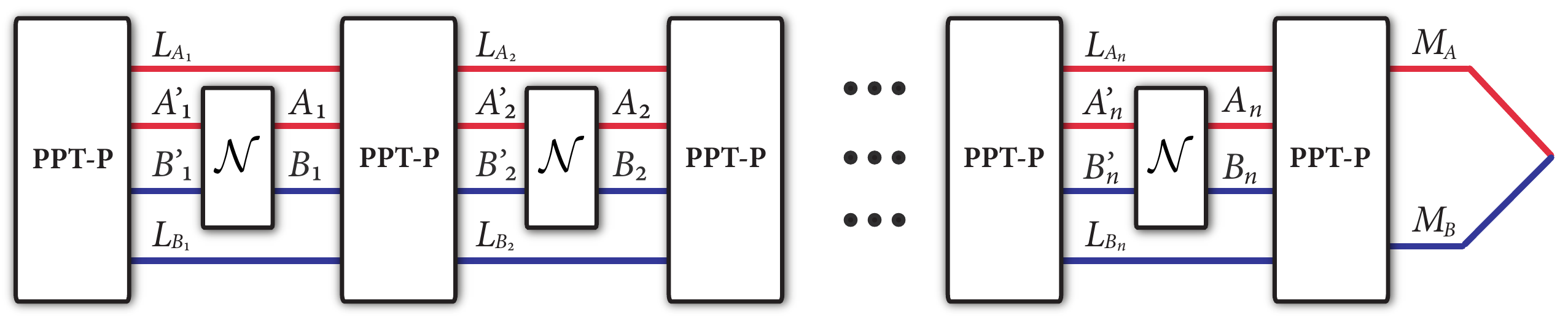}
		\caption{A model of an adaptive positive-partial-transpose (PPT) assisted entanglement generation protocol using a bidirectional channel $\mathcal{N}$. Secret-key agreement proceeds analogously, if we replace the PPT-preserving channels by LOCC channels.}%
		\label{fig1}
\end{figure}

Let us formalize what we mean by entanglement and secret key distillation via a bipartite interaction \cite{DBW17}, as depicted in  Figure~\ref{fig1}. Given a bidirectional channel $\mathcal{N}_{A'B'\to AB}$, we consider entanglement (or key) distillation  protocols as follows: an initial PPT-preserving (or LOCC) channel  between Alice and Bob creates a state $\rho^{(1)}_{L_{A_1}A'_1:B'_1L_{B_1}}$, where subsystems $L_{A_1}A'_1$ and $B'_1L_{B_1}$ are held by Alice and Bob, respectively. Note that a bipartite channel $\mathcal{P}_{A'B'\to AB}$ is PPT-preserving if
$\T_B \circ \mathcal{P}_{A'B'\to AB} \circ \T_{B'}$ is a channel \cite{Rai99,Rai01}. Furthermore, a bipartite channel is PPT-preserving if and only if its Choi operator is a PPT state \cite{Rai01}. An LOCC channel is a particular example of a PPT-preserving channel \cite{Rai99,Rai01}. The dimensions of the auxiliary systems $L_{A_1}$ and $L_{B_1}$ are finite, but can be arbitrarily large. Subsystems $A'_1$ and $B'_1$ of $\rho^{(1)}_{L_{A_1}A'_1:B'_1L_{B_1}}$ are then inserted into the channel $\mathcal{N}$, yielding a state $\sigma^{(1)}_{L_{A_1}A_1:B_1L_{B_1}}$. This is followed by $n$ more  PPT-preserving (or LOCC) channels interleaved with $n$ uses of the channel. After $n$ channel uses, the final PPT-preserving (or LOCC) channel should yield a state $\omega_{M_A M_B}$ that has fidelity \cite{U76} larger than $1-\varepsilon$ with a maximally entangled state  $\Phi_{M_AM_B}$ containing $\log_2 M$ ebits (or a private state containing $\log_2 K$ private bits 
 between Alice and Bob).  Such a protocol is called an $(n,M(\text{or }K),\varepsilon)$-protocol. A rate $R$ is achievable if for $\varepsilon \in (0,1)$, $\delta > 0$, and sufficiently large $n$, there exists an $(n,2^{n(R-\delta)},\varepsilon)$ protocol. The largest achievable rate is the PPT-assisted quantum capacity $Q^{2\to2}_\text{PPT}$ (or secret-key agreement capacity $P^{2\to2}_\text{LOCC}$) of $\mathcal{N}$.
 
 By private states containing $\log_2 K$  private bits, we mean states $\gamma_{K_AS_A:K_BS_B}$, such that measurement of the $K_{A,B}$ subsystems, the \emph{key part}, yields $\log K$  bits of secret key as long as the $S_{A,B}$ subsystems, the \emph{shield part}, are kept secure from Eve, who is allowed to be in control of the purification of $\gamma$. See the seminal works \cite{HHHO05,HHHO09} for further details.

The main results of this paper are \textit{strong converse} bounds on $Q^{2\to2}_\text{PPT}$ and $P^{2\to2}_\text{LOCC}$, in terms of the bidirectional max-Rains information and bidirectional max-relative entropy of entanglement, respectively. The strong-converse nature of the bound means that the error~$\varepsilon$ tends to one in the limit of many channel uses if the communication rate exceeds the bound. Our first result is as follows:

\begin{theorem}\label{theo1}
The PPT-assisted quantum communication capacity of a bidirectional channel $\mathcal{N}$ is bounded from above by its bidirectional max-Rains information:
$
{Q}^{2\to 2}_{\PPT}(\mathcal{N})\leq R^{2\to 2}_{\max}(\mathcal{N})$,
and this upper bound is a strong converse bound.
\end{theorem}

Theorem \ref{theo1} is a consequence of the observation that the bidirectional max-Rains information of a bidirectional channel $\mathcal{N}$ cannot be enhanced by amortization; i.e., for an input state $\rho_{L_AA'B'L_B}$, the following holds
\begin{equation}\label{amor}
R_{\max}(L_A A; B L_B)_{\mathcal{N}(\rho)}\leq R_{\max}(L_A A';B' L_B)_{\rho}+ R^{2\to 2}_{\max}(\mathcal{N}),
\end{equation}
where $R_{\max}(A;B)_{\rho} := \inf\{ \lambda : \rho_{AB} \leq 2^{\lambda} \sigma'_{AB} , \sigma'_{AB} \in \operatorname{PPT}'(A\!:\!B)\}$ denotes the max-Rains information of the state $\rho_{AB}$ \cite{TWW17}, with $\operatorname{PPT}'(A\!:\!B)$ denoting the set of all positive semidefinite operators $\sigma'_{AB}$ such that the trace norm $\Vert \T_B(\sigma'_{AB})\Vert_1\leq 1$ \cite{AdMVW02}. This observation was made in the case of point-to-point channels \cite{BW17} and constitutes a contribution of our companion paper \cite{DBW17}. By successive application of the amortization relation in \eqref{amor} to every use of $\mathcal{N}$ in an $(n,M,\varepsilon)$-protocol, it follows that $R_{\max}(M_A;M_B)_\omega\leq nR^{2\to 2}_{\max}(\mathcal{N})$, where $|M_A|=|M_B|=M$. As, by assumption, $\Tr[\Phi_{M_AM_B}\omega_{M_AM_B}]\geq1-\varepsilon$, whereas by \cite[Lemma~2]{Rai99}, $\Tr[\Phi_{M_AM_B}\sigma'_{M_AM_B}]\leq\frac{1}{M}$ for any $\sigma'_{M_AM_B}\in\operatorname{PPT}'(A\!:\!B)$, it follows by a data-processing argument that $R_{\max}(M_A;M_B)_\omega\geq\log [(1-\varepsilon)M]$. Hence we obtain
\begin{equation}\label{eq:rains-ent-dist-strong-converse}
\frac{1}{n}
\log_2M\leq R^{2\to 2}_{\max}(\mathcal{N})+\frac{1}{n}\log_2\!\left(\frac{1}{1-\varepsilon}\right),
\end{equation}
which implies Theorem~\ref{theo1}. Solving \eqref{eq:rains-ent-dist-strong-converse} for $\varepsilon$ shows that the error increases exponentially fast to one if the rate exceeds $R^{2\to 2}_{\max}(\mathcal{N})$, establishing the strong converse nature of the bound. 

As an example, we have numerically computed $R^{2\to 2}_{\max}$ for the qubit partial swap operation \cite{FHSSW11,audenaert2016entropy}, which is performed by application of the unitary $U_p=\sqrt{p}I+\iota\sqrt{1-p}S$, where $S=\sum_{ij}|ij\rangle\langle ji|$ is the swap operator. Such an operation can be compared to a beamsplitter \cite{konig2013limits}. We also consider when the partial swap is followed by a traceout of Alice's subsystem. As another example, we have computed  $R^{2\to 2}_{\max}$ for a qubit swap operator with collective dephasing \cite{palma1996quantum}, which is a typical model for noise in a quantum computer. In the qubit case, a collective phase rotation acts as $|0\rangle\to|0\rangle$, $|1\rangle\to e^{\iota\phi}|1\rangle$ for some phase $\phi$. Hence $|00\rangle\to|00\rangle$,  $|01\rangle\to e^{\iota\phi}|01\rangle$, $|10\rangle\to e^{\iota\phi}|10\rangle$, and $|11\rangle\to e^{2\iota\phi}|11\rangle$. The collective phase rotation occurs with probability $1-p$. 

\begin{figure}
		\centering
\includegraphics[width=0.7\linewidth, trim={1cm 5cm 1cm 8cm}]{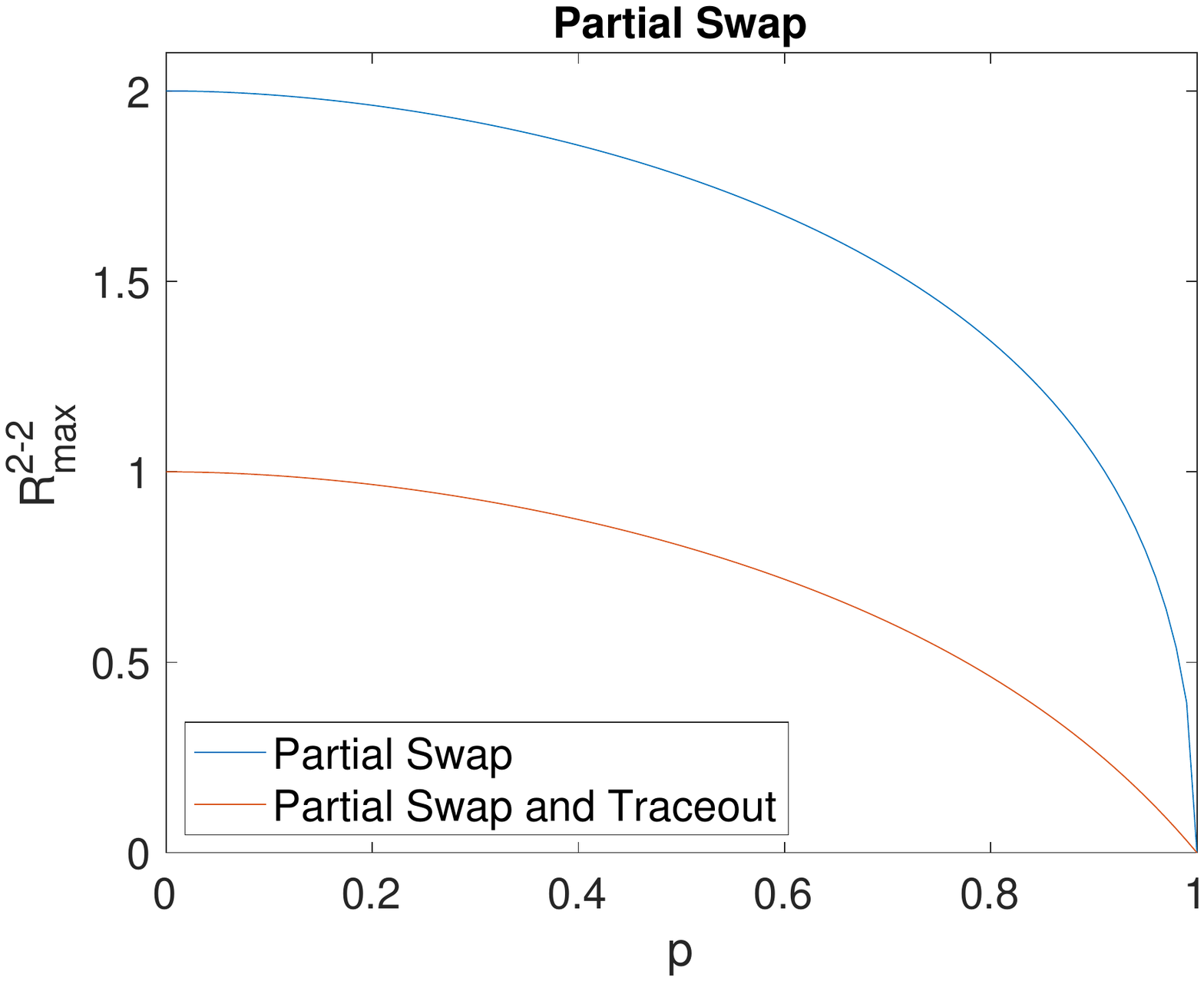}
		\includegraphics[width=0.7\linewidth, trim={1cm 7.5cm 1cm 8cm}]{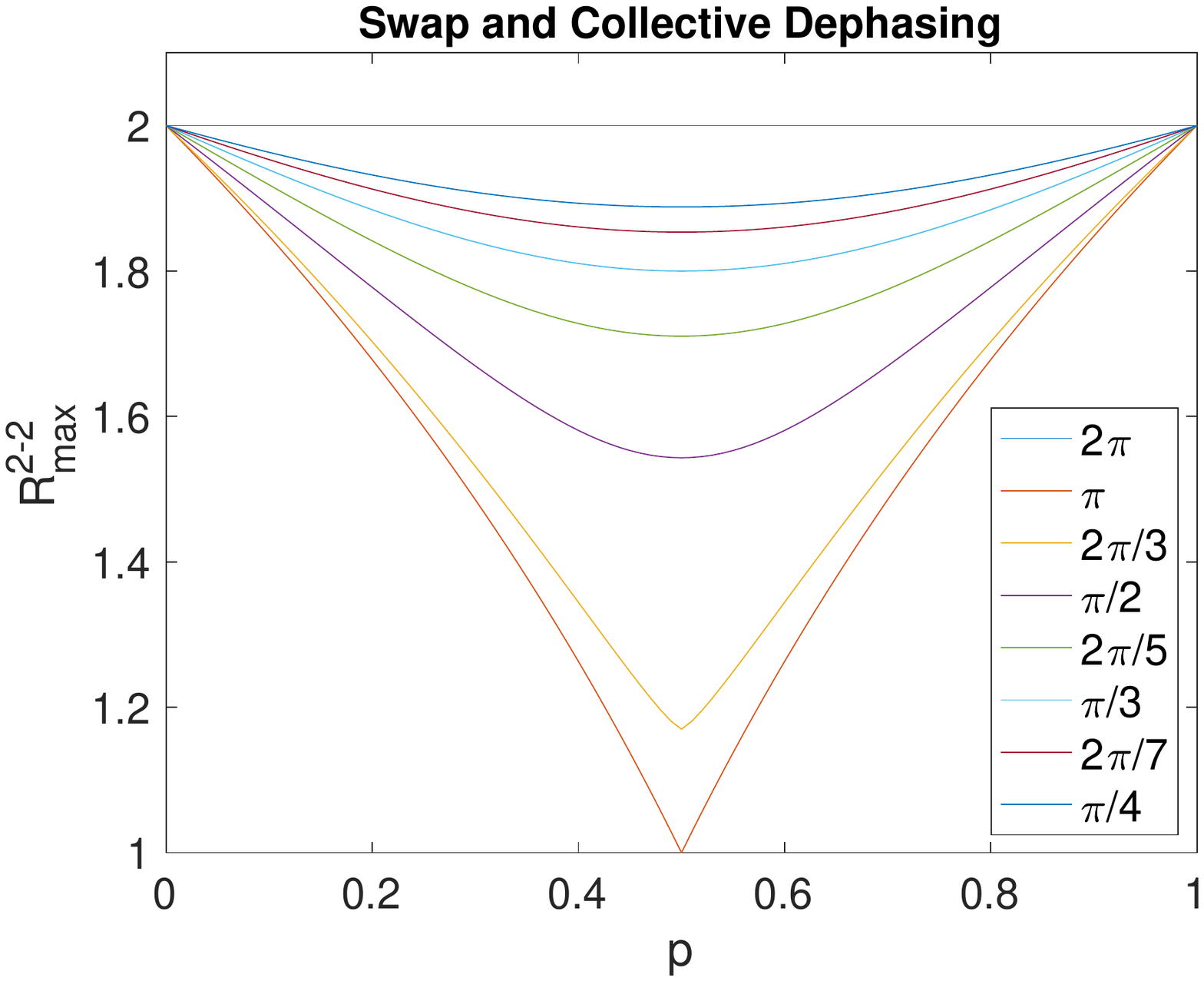}
		\caption{Our bounds plotted versus the channel parameter $p$. From top to bottom, they are (i) qubit partial swap operation and qubit partial swap operation followed by traceout of Alice's output and (ii) a qubit swap operation with collective dephasing for various phases $\phi$.}
		\label{fig2}
\end{figure}

Our results are plotted in Figure \ref{fig2}. For the partial swap, the top plot shows the expected decline from two ebits to zero, as the channel tends towards total depolarization. For the partial swap and traceout, the decline is from one ebit to zero. In the example of collective dephasing, as expected, the performance is the worst at $p=1/2$, where there is the most uncertainty about whether the collective phase rotation has taken place. For  $\phi=\pi$, we can have a reduction of a factor of $1/2$. Let us remark that this bound can actually be achieved. To do so, Alice and Bob both locally create two Bell states $\Phi^+_{L_AA'}$ and $\Phi^+_{B'L_B}$, which are maximally entangled. After the swap operation and the collective dephasing, they end up sharing the state $\frac{1}{2}\Phi^{+}_{AL_B}\otimes\Phi^{+}_{BL_A}+\frac{1}{2}\Phi^{-}_{AL_B}\otimes\Phi^{-}_{BL_A}$. To find out the phase, Alice and Bob can locally measure either $A$ and $L_B$ or $L_A$ and $B$ in the Pauli-$X$ basis, thus sacrificing one ebit. If their results agree, they have $\Phi^{+}$, and otherwise $\Phi^{-}$, which can be rotated to~$\Phi^{+}$ via local unitary.

For the generation of secret key, we have the following:
\begin{theorem}\label{theo2}
The secret-key agreement capacity of a bidirectional channel $\mathcal{N}$ is bounded from above by its bidirectional max-relative entropy of entanglement:
$
{P}^{2\to 2}_{\LOCC}(\mathcal{N})\leq E^{2\to 2}_{\max}(\mathcal{N})$,
and this upper bound is a strong converse bound.
\end{theorem}
Theorem \ref{theo2} is a consequence of the amortization property of the bidirectional max-relative entropy of entanglement, which follows from the \emph{data processed triangle inequality} for the max-relative entropy of entanglement \cite{CM17}. The proof then follows along the lines of that for Theorem~\ref{theo1}, while making use of the relation between tripartite key states and bipartite private states and the privacy test from \cite{WTB16}.

If a bidirectional channel has certain symmetries, tighter bounds than the ones given in Theorems \ref{theo1} and \ref{theo2} can be obtained: A bidirectional channel $\mathcal{N}_{A'B'\to AB}$ is said to be PPT-simulable  (or teleportation-simulable \cite{STM11})   with associated resource state $\theta_{D_AD_B}$, for some auxiliary quantum systems $D_A$ and $D_B$,  if there exists a PPT-preserving (or LOCC) channel $\mathcal{P}_{D_AA'B'D_B\to AB}$ such that
$
\mathcal{N}_{A'B'\to AB}\left(\rho_{A'B'}\right)=\mathcal{P}_{D_AA'B'D_B\to AB}\left(\rho_{A'B'}\otimes\theta_{D_AD_B}\right). 
$
If a bidirectional channel is PPT-simulable (or teleportation-simulable), then the bounds given in Theorem \ref{theo1} (or Theorem \ref{theo2}) reduce to the standard Rains relative entropy \cite{Rai99} (or the relative entropy of entanglement \cite{vedral1998entanglement}) of the resource state.

In particular, it can be shown that any bicovariant bidirectional channel is teleportation-simulable, hence also PPT-simulable, with the normalized Choi state as the associated resource state. By bicovariant, we mean that for finite groups $G$ and $H$, with representations as unitary one-designs,  the following holds for all $g\in G$, $h\in H$ and all input states $\rho_{A'B'}$:
$\mathcal{N}_{A^{\prime}B^{\prime}\rightarrow AB}((\mathcal{U}_{A^{\prime}
 }(g)\otimes\mathcal{V}_{B^{\prime}}(h))(\rho_{A^{\prime}B^{\prime} }))=(\mathcal{W}_{A}(g,h)\otimes\mathcal{T}_{B}(g,h))(\mathcal{N}_{A^{\prime}B^{\prime}\rightarrow AB}(\rho_{A^{\prime}B^{\prime}}))$,
 for unitary representations $g\rightarrow U_{A^{\prime}}(g)$, $h\rightarrow V_{B^{\prime}}(h)$, $(g,h)\rightarrow W_{A}(g,h)$ and $(g,h)\rightarrow T_{B}(g,h)$, where we have defined $\mathcal{U}(g)(\cdot):= U(g)(\cdot)\left(U(g)\right)^{\dag}$. An example of a bicovariant channel is the CNOT gate \cite{G99,GC99}, or one that applies the CNOT gate with some probability and replaces with the maximally mixed state with the complementary probability. 

\textit{Private Reading}---Consider the task of reading a message stored in a memory device, while under the surveillance of a passive eavesdropper Eve. The read-out of the stored message should be private, under the assumption that Eve has complete access to the environment but no direct access to the device. Such a private reading protocol is a private version of the quantum reading protocol from \cite{DW17} (see also \cite{BRV00,Pir11}). Formally, in a private reading protocol, the encoder, Alice, encodes a secret classical message $k\in\mathcal{K}$ into a sequence of wiretap channels chosen from a set $\mathcal{M}_{\mathcal{X}}:=\{\mathcal{N}^x_{B'\to BE}\}_{x\in\mathcal{X}}$, by means of codewords $x^n(k)=x_1(k)\cdots x_n(k)$. We call the set of wiretap channels a \emph{wiretap memory cell}, where the dimensions of the systems $B'$, $B$, and $E$ are independent of~$x$. It is assumed that Eve   has access to the $E$ systems only, but her computational power may be unbounded. As a special case, we can consider isometric memory cells, which map the input space $B'$ reversibly into the output space $BE$. The memory device containing the channels is then delivered to the reader, Bob, as a read-only device. 

Bob can use quantum inputs, channels, and  measurements to read out the message encoded in the device. In particular, he can apply an adaptive strategy consisting of creating an initial state $\rho^{(1)}_{B'_1S_{B_1}}$, inserting $B'_1$ into the channel $\mathcal{N}^{x_1}$, applying a quantum channel on the output $B_1L_{B_1}$, which results in a new state $\rho^{(2)}_{B'_2L_{B_2}}$, the $B'_2$ subsystem of which is then entered into $\mathcal{N}^{x_2}$ and so on. After using  all $n$ channels, interleaved by quantum channels, Bob then performs a final measurement, yielding an estimate $\hat{k}$ of the encoded message.

As mentioned above, the channels are wiretapped by an eavesdropper Eve. As is the case for Bob, the device is assumed to be read-only for Eve as well. So she assumes the role of a passive eavesdropper and only has access to the output systems $E_1,\ldots, E_n$ of the channels $\mathcal{N}^{x_1},\ldots,\mathcal{N}^{x_n}$, respectively. The goal is to maximize Bob's success probability of guessing the message, while restricting Eve to obtain negligible information about the message. 

In the case of an isometric wiretap memory cell $\mathcal{M}_{\mathcal{X}}=\{\mathcal{U}^x_{B'\to BE}\}_{x\in\mathcal{X}}$, Theorem \ref{theo2} provides a (strong converse) upper bound on the maximum achievable rate of a private reading protocol. This follows from the observation that in a \emph{purified} setting \cite{HHHO05,HHHO09,WTB16}, in which  purifications of all input states are considered and for every operation the ancillary subsystems are being considered as well, a private reading protocol can be used to create a private state, containing $K=|\mathcal{K}|$ bits of secret key, between Alice and Bob. To do so, Alice prepares a purification
$\frac{1}{\sqrt{K}}\sum_{k\in\mathcal{K}}|k,k,k\rangle_{K_A\hat{K}C}$ of a maximally classically correlated state $\frac{1}{K}\sum_{k\in\mathcal{K}}|k,k\rangle\langle k,k|_{K_AC}$ and encodes subsystem $C$ by means of an isometry $|k\rangle_C\to|x^n(k)\rangle_{X^n}$. For every letter $x_i(k)$ of the codeword, the combined operation of Alice's writing and Bob's readout of the memory device is then described by a controlled isometry
\begin{equation}\label{eq:iso-v-m}
U^{\mathcal{M}_\mathcal{X}}_{X_iB'_i\to X_iB_iE_i}:= \sum_{x\in\mathcal{X}}|{x}\rangle \langle{x}|_{X_i}\otimes{U}^{x}_{B'_i\to B_iE_i}.
\end{equation} 
In an adaptive protocol, the $U_i$'s are interleaved with Bob's operations. This is then followed by a decoding channel on Bob's side, after which Alice and Bob's state should be $\varepsilon$-close to a private state $\gamma_{K_AS_A:K_BS_B}$, where $S_A$ and $S_B$ denote the shield parts containing all ancillary systems that Alice and Bob have created during the purified protocol (see \cite[Section~6.3]{DBW17}). Defining a bidirectional channel $\mathcal{N}^{\mathcal{M}_\mathcal{X}}_{XB'\to XB}(\cdot):=\Tr_{E}[U^{\mathcal{M}_\mathcal{X}}_{XB'\to XBE}(\cdot)(U^{\mathcal{M}_\mathcal{X}}_{XB'\to XBE})^\dagger]$, it is straightforward to conclude that the purified reading protocol is an example of a  bidirectional secret-key-agreement protocol. Hence by Theorem~\ref{theo2}, its capacity is  bounded from above by $E^{2\to2}_{\max}(\mathcal{N}^{\mathcal{M}_\mathcal{X}}_{XB'\to XB})$.

As a concrete example, let us consider a \emph{qudit erasure wiretap memory cell} \cite{DW17}. It is defined as $\bar{\mathcal{Q}}^p_{\mathcal{X}}=\{\mathcal{Q}^{p,x}_{B'\to BE}\}_{x\in\mathcal{X}}$, where $\mathcal{Q}^{p,x}(\cdot)=U^p\sigma^x(\cdot)\left(\sigma^x\right)^\dagger\left(U^p\right)^\dagger$, with Heisenberg--Weyl operators $\sigma^x$ and
$
U^p \vert \psi\rangle_{B'}= \sqrt{1-p}\vert \psi \rangle_B \vert e \rangle_E + \sqrt{p}|e\rangle_B \vert \psi\rangle_E$.
is the isometric extension of the erasure channel. Using a covariance argument, we reduce the upper bound in Theorem~\ref{theo2} to the relative entropy of entanglement of the Choi state, which provides a strong converse upper bound of $2(1-p)\log_2 d$ on the private reading capacity of $\bar{\mathcal{Q}}^p_{\mathcal{X}}$.

\textit{Summary and Outlook}---We have provided strong converse upper bounds on the PPT-assisted quantum capacity and the LOCC-assisted private capacity of a bidirectional quantum channel. The bound on the quantum capacity is related to the Rains bound \cite{Rai99,Rai01}, as well as that in \cite{WFD17}, and can be efficiently computed by SDP solvers. We have provided examples that demonstrate the applicability of our bound. The bound on the private capacity is in terms of the max-relative entropy of entanglement \cite{D09,Dat09,CM17}. As an application, we have considered the task of private reading in the presence of a passive eavesdropper. Both bounds can be improved in the case of a bicovariant bidirectional channel. As an example, we have upper bounded the private reading capacity of a qudit erasure wiretap memory cell. 
Future directions from here include  generalising our results from bi- to multipartite quantum interactions, which could be effectively applied in the theory of quantum networks.

\begin{acknowledgments}
We thank K.~Azuma, A.~Harrow, M.~Huber, C.~Lupo, B.~Munro, M.~Murao, and G.~Siopsis for discussions. SD acknowledges support from
the LSU Graduate School Economic Development Assistantship and the LSU Coates Conference Travel Award. MMW acknowledges support from the US Office of Naval Research and the National Science Foundation. Part of this work was completed during the workshop \textquotedblleft Beyond i.i.d.~in Information Theory,\textquotedblright\ hosted by the Institute for Mathematical Sciences, NUS Singapore,
24-28 July 2017.
\end{acknowledgments}


\end{document}